# Disentangling Internal and Forced Climate Variability with Convolutional Neural Networks using Multivariate Fields

Guillaume Gastineau [a], Elena Provenzano [a], Constantin Bône [a],

Nicolas Lebas [a] and Robert C. J. Wills [b].

[a] *UMR LOCEAN, IPSL, Sorbonne Université, IRD, CNRS, MNHN, Paris, France*

[b] *Institute for Atmospheric and Climate Science, ETH Zurich, Switzerland*

*Corresponding author*: Guillaume Gastineau (guillaume.gastineau@locean.ipsl.fr)

Submitted to *Artificial Intelligence for the Earth Systems* August 24th 2026

ABSTRACT

Long-term climate data exhibit variations composed of internal and forced components. Internal variability arises from natural processes that could occur within a stable climate. Forced variability, on the other hand, reflects climate changes induced, for example, by anthropogenic greenhouse gas and aerosol emissions. Accurately distinguishing between these types of variability is crucial for attributing climate fluctuations and understanding internal variability processes and impacts. In this study, we apply a U-Net convolutional network, a model commonly used in computer vision, to separate forced and internal climate variability from 1950 to 2022 using a multi-model dataset. The dataset includes multiple fields from five single-model initial-condition large ensembles. Cross-validation is conducted by training the U-Net using the data from the ensembles of four models, leaving out the data from one model to assess performance. Validation results yield errors ranging from 0.1°C to 0.4°C for the forced variability of local-scale monthly surface air temperature, which is no more than half the magnitude of external forcing. The U-Net achieves better performance than a simple approach based on a fourth-order polynomial trend for estimating forced variability. The error is mainly due to insufficient sampling and poor agreement among the models, as the U-Net underestimates the warming for the model with the highest transient climate sensitivity during validation. Performance is lower for variables like sea-level pressure and precipitation because of their low ratio of forced to internal variability. This framework might be enhanced by incorporating a larger multi-model dataset in the training and validation.

SIGNIFICANCE STATEMENT

Climate variations can be decomposed into internal fluctuations and externally forced changes. The internal fluctuations represent naturally occurring processes due to atmospheric and oceanic flow. The forced changes are mainly due to anthropogenic emissions and include the global warming tendency. Here, we design a method to distinguish internal and forced variability based on a U-Net, a standard approach used in computer vision. The U-Net is trained using data from various climate models and uses multiple fields as input, such as temperature, sea-level pressure, or precipitation. Evaluation reveals that the method manages to attribute most of the forced temperature variations. Using a larger dataset could improve these results and lead to a more accurate assessment of climate variability processes.

## 1. Introduction

Forced climate variability designates climate fluctuations due to the evolution of external forcings. For instance, the rising atmospheric concentration of greenhouse gases is responsible for the warming of air temperatures over the last century (Stott et al., 2010, Ribes and Terray, 2013; Eyring et al., 2021; Gillett et al., 2021). The changes in anthropogenic aerosols have significantly attenuated this warming, although major uncertainties remain on the amplitude and timing of this effect (Moseid et al., 2020; Dittus et al., 2020; Eyring et al., 2021; Kalisoras et al. 2024). Additionally, the climate is also modulated by forcing with a natural origin, such as the ones originating from the injection of volcanic aerosols or the variations in the incoming solar radiation. Although uncertainties remain regarding these natural forcings, their large-scale effect is estimated to be much smaller than anthropogenic forcings during 1880-2022, except for the few years following the three major volcanic eruptions of Agung in 1963, El Chichón in 1982, and Pinatubo in 1991 (Stenchikov et al., 1998; Stott et al., 2000; Menegoz et al., 2018). Lastly, the ozone in the stratosphere (Shepherd, 2008; Revell et al. 2022), as well as land-use (Luo et al., 2024) changes, also modify the climate, although their contributions are weaker than those of greenhouse gases or aerosols.

Internal variability designates the climate variability resulting from the chaotic nature of the flow within the atmosphere and ocean. Internal variability is composed of specific modes of climate variability, with El Niño Southern Oscillation (ENSO) being the most important mode at the global scale (Jin et al., 1997; Wang and Picaut, 2004). Other important modes of variability at the decadal to multi-decadal time scales are the Atlantic Multidecadal Variability (AMV; Knight et al. 2006) and the Pacific Decadal Oscillation (PDO; Mantua et al., 1997; Newman et al., 2016). The Southern Ocean also shows some important variability, with the Southern Annular Mode in the atmosphere (Thompson and Wallace, 2000) and low-frequency fluctuations in the ocean (Zhang et al., 2019; Hogg and Blundell, 2006; Le Bars et al., 2016). Such internal variability can be modulated by the mean climate state, as a warming climate may lead to more intense ENSO events (Maher et al., 2023; Cai et al., 2023). The decadal fluctuations of the sea surface temperature in the Atlantic Ocean might also decrease in magnitude as the climate warms, and as the Atlantic Meridional Overturning Circulation variability diminishes in climate models projections (Cheng et al., 2016; Ma et al., 2021; Bonnet et al., 2021).

The separation of internal and forced variability is equivalent to attributing and quantifying the climate fluctuations due to internal ocean-ice-atmosphere processes and those due to external forcing changes. This is a notably difficult task, as internal and external variability often share the same spatial pattern (Deser and Philips, 2023) and can have similar time evolution. However, distinguishing the internal and forced variability is key to trustfully assess the role of external forcing at the regional scale. This separation could also help to better understand the intrinsic uncertainty in climate projections due to internal variability (Hawkins and Sutton 2009; Lehner et al. 2020). This will also allow a more accurate estimate of the effects of external forcing, which can have effects on variability in addition to the secular warming (Bellomo et al. 2018; He et al., 2023). Among other examples, anthropogenic aerosols (Booth et al., 2012; Bellucci et al. 2017; Undorf et al. 2018, Dittus et al., 2021) and natural forcing (Borchert et al., 2021) were found to explain much of the past SST variability in the North Atlantic Ocean. The recent persistent cooling in the tropical Pacific, observed over the last three decades, is also debated, as it is not reproduced in climate models (Kociuba and Power, 2015; Watanabe et al., 2021, Wills et al. 2022). Possible explanations invoke a potential role of internal variability poorly simulated in models (Kociuba and Power, 2015), tropical Pacific mean-state biases (Seager et al. 2019), or a missing effect of external forcing on the Southern Ocean evolution (Hartmann 2022; Dong et al. 2022; Kang et al., 2023). A precise separation of the internal and forced variability would also be beneficial for variables with large uncertainties in model projections and large societal impacts, such as precipitation in monsoon regions (Seth et al., 2019) or the summer surface air temperature over western Europe (Vautard et al., 2023). Lastly, a separation between forced and internal variability might also improve our understanding of the internal modes of variability, which can be affected by the method used to remove the external forcing effect, as seen in the case of the AMV (Frankcombe et al. 2015; Qasmi et al., 2017; Deser and Philips, 2023).

Several methods have been previously introduced for the separation of internal and forced variability. Among the methods routinely used, the effect of external forcing can be estimated by a trend, or by a regression onto the global mean surface temperature time series. However, such methods can lead to important errors, especially regarding the estimated multidecadal changes (Deser and Phillips, 2023). Methods such as a linear inverse modeling filter (Frankignoul et al., 2017) or low-frequency component analysis filtering (Wills et al., 2020) have been shown to perform better than using trends or regression when using data produced by single model initial condition large ensembles. However, there has been a lack of systematic comparison of these methods using the same benchmark data. Moreover, deep learning methods have not yet been extensively applied to separate internal and forced variability, despite some promising previous attempts (Sweeney et al., 2023; Bône et al., 2024). Recently, the Forced Component Statistical Method Intercomparison Project (ForceSMIP; https://sites.google.com/ethz.ch/forcesmip/; Wills et al. 2026) was designed to comprehensively assess these methods using data from single-model large ensemble simulations (Deser et al., 2020). Using these large ensembles allows estimating the forced variability for each climate model through the averaging of several ensemble members starting from different initial conditions. The error in separating the internal and forced variability can then be estimated by comparing the result of each method to the ensemble mean of each model.

In this manuscript, we present a deep learning statistical model designed to separate the forced and internal variability. This model is trained and tested using the ForceSMIP framework (Wills et al., 2026) and following Bône et al. (2024), with a convolutional neural network. Here, many improvements have been implemented. First of all, the method used here relies on the use of multiple monthly fields. Second, the method originally used a noise-to-noise approach, while we use here a more classical supervised learning method. Lastly, the method presented is also more extensively evaluated and optimized.

In the second section, the data are presented, and the deep learning model is described. The third section is dedicated to an overview of the performance of the model. The last section is dedicated to discussions and conclusions.

## 2. Data and Methods

### *a. Data*

We use the ForceSMIP Tier1 data (Wills et al. 2025; 2026). It includes a training dataset with outputs from five large ensembles of climate model simulations (see Table S1): CESM2, CanESM5, MIROC-ES2L, MIROC6 and MPI-ESM-1-2-LR. The simulation period is from 1880 to 2100. All simulation ensembles are built by integrating multiple members starting from different equiprobable initial conditions. All simulations used the Coupled Model Intercomparison Project Phase 6 (CMIP6) historical external forcing from 1880 to 2014, and scenario forcings after 2015. CESM2, CanESM5, MIROC-ES2L, MIROC6 and MPI-ESM-1-2-LR used SSP3-7.0, SSP5-8.5, SSP2-4.5, SSP5-8.5 and SSP5-8.5, respectively.

We use the surface air temperature (*tas*), the sea surface temperature (SST, or *tos*), the zonal-mean air temperature (*zmta*), the sea-level pressure (SLP, or *psl*), the precipitation (*pr*), the monthly maximum of daily precipitation (*monmaxpr*), the monthly maximum of daily temperature maximum (TXx, or *monmaxtasmax*), and the monthly minimum of daily temperature minimum (TNn, or *monmintasmin*).

The observational data is ERSST version 5 for the SST (Huang et al., 2017) and ERA5 (Hersbach et al., 2020) for the other variables, and is extracted in 1950-2022.

All data are interpolated onto a common longitude-latitude 2.5°x2.5° horizontal grid, except for the zonal-mean air temperature, where an interpolation on a 2.5° latitude grid and 17 vertical standard pressure levels (from 1000 hPa to 10 hPa) was performed.

*b. Preprocessing*

First, the anomalies are calculated by subtracting a climatology from the 1950-2022 period. For each of the five training models, we then compute the ensemble mean of each field. Hereafter, $X_{ij}$ indicates the raw anomalies of field $X$ from model $j$ ($1 \leq j \leq 5$) and member $i$ ($1 \leq i \leq n_j$, with $n_j$ the ensemble size). The ensemble mean for the model $j$, called $\overline{X_j}$, estimates the forced variability. The deviations from the ensemble mean, $X'_{ij}$=$X_{ij}$-$\overline{X_j}$, estimate the internal variability.

The data were normalized before training. First, we calculate the global mean of the time standard deviation of $X_{ij}$, $\overline{X_j}$ and $X'_{ij}$, which provides $\sigma_{X,i,j}$, $\sigma_{\overline{X},j}$ and $\sigma_{X',i,j}$. We calculate the ensemble mean values of $\sigma_{X,i,j}$ and $\sigma_{X',i,j}$ and then calculate an average for all models used in the training, which provides three scalars $\overline{\sigma}_{\overline{X}}$, $\overline{\sigma}_X$and $\overline{\sigma}_{X'}$. The anomalies from $X_{ij}$, $\overline{X_j}$ and $X'_{ij}$ are then normalized by dividing by $\overline{\sigma}_X$, $\overline{\sigma_{\overline{X}}}$ and $\overline{\sigma}_{X'}$, respectively.

All variables but the zonal-mean temperature share the same latitude and longitude coordinates. To treat several variables together, the zonal-mean temperature field with latitude and pressure level as coordinates (see for instance Fig. S1a) is mapped into equivalent longitude and latitude coordinates. First, the pressure levels are duplicated according to the pressure thickness of each layer to transform the vertical grid from 17 levels to 72 levels, from 1000 hPa to 10 hPa. Then, the obtained 72-level field is concatenated to a vertically flipped copy of the same field (which then goes from top 10 hPa to 1000 hPa). The resulting field is circular along the vertical axis and is analogous to a longitude axis (see Fig. S1b).

*c. Baseline model*

A fourth-order polynomial trend is calculated at each grid point for each variable. As in Wills et al. (2026), the polynomial trend provides the monthly estimate of the forced changes. Hereafter, the trend line provides a baseline model that will be compared to the U-Net estimates.

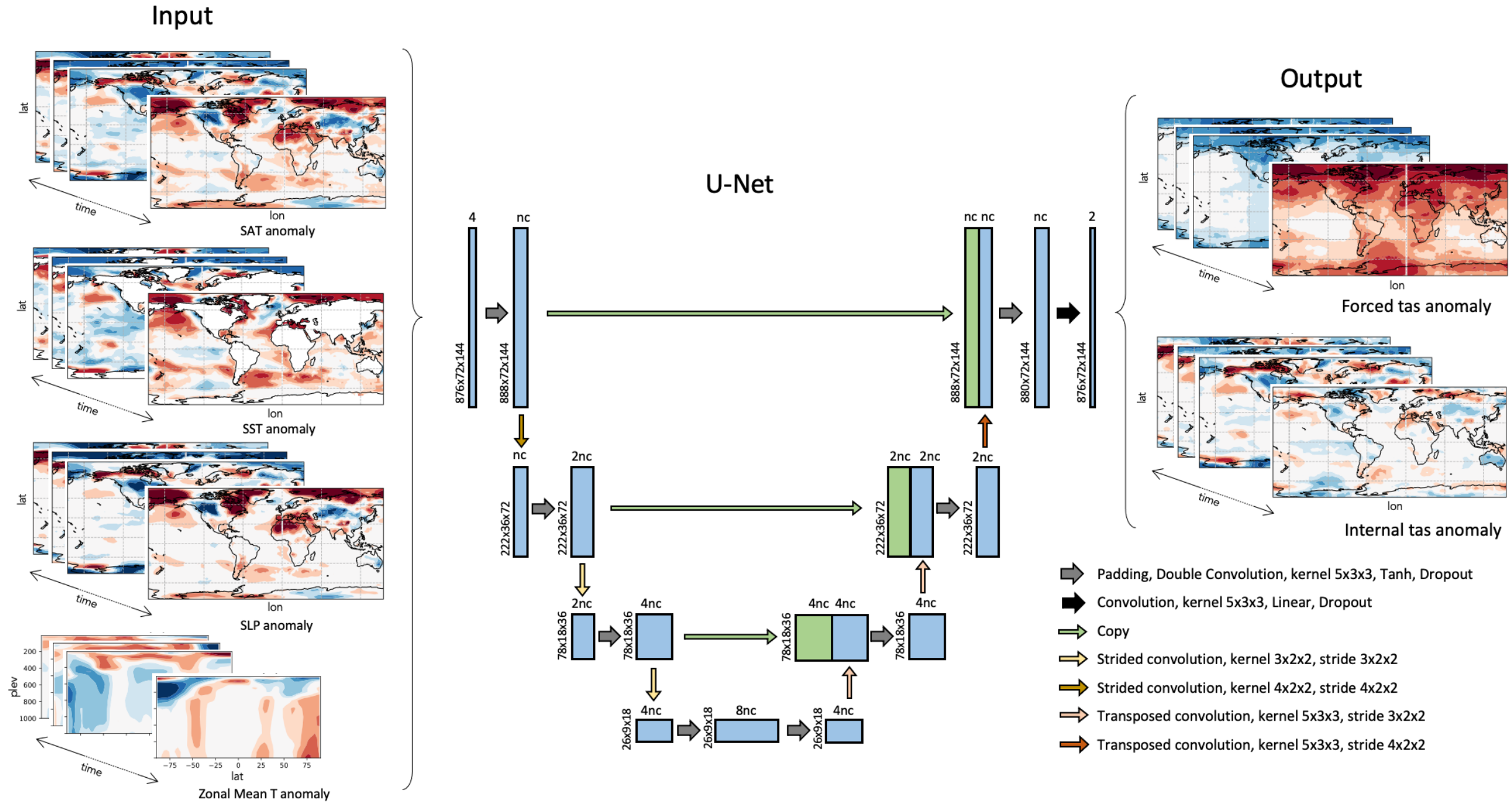

Fig. 1. Schematic of the U-Net architecture with the associated input and output, for the case of U-Net_tas dedicated to the separation of the internal and forced variability for the surface air temperature (SAT).

*d. U-Net Convolutional Neural Network*

A U-Net convolutional neural network is designed to separate the forced and internal variability. The U-Nets designed to process *tas*, *tos*, *psl* and *zmta* have the same inputs (see Table 1), but have different outputs. They use as inputs ($X_{ij}$) the monthly normalized anomalies from the surface air temperature (*tas*), SST (*tos*), SLP (*psl*), and transformed zonal-mean temperature (*zmta*), which are provided to the U-Net as four channels, each with the dimension (144,72,876). The output is made of two channels using the same dimension, with the normalized ensemble mean ($\overline{X_j}$) and deviations from the ensemble mean ($X'_{ij}$) of the field of interest, i.e., the forced and internal component of variations. We train a U-Net for each of the variables used as inputs.  (see Table 1). Attempts using a one-channel output with only the normalized ensemble mean, $\overline{X_j}$, systematically show less skill than when using the two channels, as found by Sweeney et al. (2023). We speculate that using these two channels in the output introduces more variability in the output data, which helps the training of the neural network.

| Model | Input | Output |
|---|---|---|
| U-Net_tas | $tas_{ij}, psl_{ij}, tos_{ij}, zmta_{ij}$ | $tas'_{ij}, \overline{tas_j}$ |
| U-Net_psl | $tas_{ij}, psl_{ij}, tos_{ij}, zmta_{ij}$ | $psl'_{ij}, \overline{psl_j}$ |
| U-Net_zmta | $tas_{ij}, psl_{ij}, tos_{ij}, zmta_{ij}$ | $zmta'_{ij}, \overline{zmta_j}$ |
| U-Net_tos | $tas_{ij}, psl_{ij}, tos_{ij}, zmta_{ij}$ | $tos'_{ij}, \overline{tos_j}$ |
| U-Net_pr | $tas_{ij}, pr_{ij}, tos_{ij}, zmta_{ij}$ | $pr'_{ij}, \overline{pr_j}$ |
| U-Net_monmaxpr | $tas_{ij}, monmaxpr_{ij}, tos_{ij}, zmta_{ij}$ | $monmaxpr'_{ij}$, $\overline{monmaxpr_j}$ |
| U-Net_monmaxtasmax | $tas_{ij}, monmaxtasmax_{ij}, tos_{ij}, zmta_{ij}$ | $monmaxtasmax'_{ij}$, $\overline{monmaxtasmax_j}$ |
| U-Net_monmintasmin | $tas_{ij}, monmintasmin_{ij}, tos_{ij}, zmta_{ij}$ | $monmintasmin'_{ij}$, $\overline{monmintasmin_j}$ |

Table 1. Name of the U-Net trained, and corresponding input and output used during the training process.

The U-Nets designed to process the precipitation (*pr*), the monthly maximum precipitation (*monmaxpr*), the monthly TXx (*monmaxtasmax*) and the monthly TNn (*monmintasmin*) are all identical to U-Net_psl, but use precipitation, monthly maximum precipitation, TXx or TNn, respectively, in place of the SLP (see Table 1).

The U-Net includes a contracting path (left, Fig. 1) to detect large-scale features with an evolution visible at the interannual time scale, reducing the input dimension from (144,72,876) to (18,9,26) at the bottleneck. An expansive path (right, Fig. 1) remaps these features to the output dimension, using skip connections (green arrows, Fig. 1) to preserve small-scale details. Here, the U-Net uses three-dimensional kernels in the longitude, latitude and time dimensions, so that the features learned are spatio-temporal variations analogous to videos. The U-Net architecture is derived from that used in Bône et al. (2024), with some notable modifications. Here, we use 3x3x5 kernels with a larger size in the time dimension than in the longitude or latitude dimension. We also apply a larger reduction in the time dimension, which has the largest size. Strided convolution is used to achieve the dimension reduction. The padding was adapted to geophysical data, with circular padding in the longitude dimension and reflective padding along the latitude dimension. Zero padding is applied for the time dimension mostly during the contracting path, with no padding in the last convolutions, to improve the estimation of endpoints. Lastly, dropout layers are added to avoid overfitting. The U-Nets contain 2 742 496 parameters. The hyperparameters used and the optimization details are provided in Text S1.

In the following, we will discuss results obtained with a preliminary version of the U-Net that was used in Wills et al. (2026), where it is referred to as UNet3D-LOCEAN. The U-Net investigated here and the one used in Wills et al. (2026) employ slightly different architectures and hyperparameters (see Text S2 for details), and we will show below that these differences result in better performance. Among the main differences, the U-Net presented here uses strided convolution instead of maxpool for dimension reduction, tanh instead of ReLU as activation functions, and a different optimizer.

*e. Training and validation*

The data for the supervised training is made of input and output pairs, with the full anomaly for the input, and the ensemble mean and the deviations from the ensemble mean for the corresponding member and period for the output. In order to train the model, we use the data from four of the five models. The data over the 1880-2100 period is typically resampled into 73-year overlapping periods starting every year from 1925 to 1975 (i.e. 1925-1997, 1926-1998, …, 1975-2057; see Text S1). To further increase the variety in the training data, a random shift in the longitude dimension is applied consistently in both input and output data for 50% of the training dataset. Validation is performed with data from 1950-2022 for each of the members of the left-out model. The later period is used because it is the target period for the evaluation in ForceSMIP. The training and validation process is conducted using each of the five models successively for validation. Unless stated otherwise, the validation score illustrated next is computed with a mean across the results obtained using each of the five models left out in the training. The error calculated is the root of the globally averaged (using area weighting) time-mean squared difference between the de-normalized forced anomaly reconstructed by the U-Net and the ensemble mean. It is referred to later as the RMSE (root mean square error) of the forced anomaly.

The U-Net_tas was also trained using a noise-to-noise approach as in Bône et al. (2024) instead of the supervised setting described above. We provide successively as input the field obtained from one ensemble member ($X_{ij}$) and as output the same field obtained from another ensemble member ($X_{kj}$ with $k \neq i$) of the same model. The noise-to-noise approach is known as an alternative method to learn the mean of the distribution of an object investigated, acting as a denoiser (Lehtinen et al., 2018). It was applied by Bône et al. (2024) using an analogy between the internal variability and noise. However, we found a slower convergence and a larger RMSE compared to the supervised procedure adopted here (not shown).

The inclusion in the training data of periods outside 1950-2022 is found to increase the skill of the U-Net_tas. However, the global mean RMSE increases again when data with very different external forcings are included (see Fig. S2). Using data from 1925-2057 for the training provides the lowest validation RMSE for all of the variables, except for the precipitation (*pr* and *monmaxpr*) where 1915-2067 leads to the lowest RMSE.

The training uses all members, with one *vote* for one *member*, with models having more members (see Table S1) being more used in the training. We tested a balanced setting, with one *vote* for each *model*, with a randomly selected set of members for the same size for each model and epoch. However, the validation RMSE is then slightly larger (not shown).

*f. Inference*

The ultimate goal of the U-Net once trained is to infer the forced variability from some unseen data. Hereafter, we illustrate the results obtained when using observations and reanalyses for the inference. We derive a set of five estimates to sample the methodological uncertainties by using the five U-Nets trained by successively excluding one of the five training models. The estimated forced variability is the average of the five estimates.

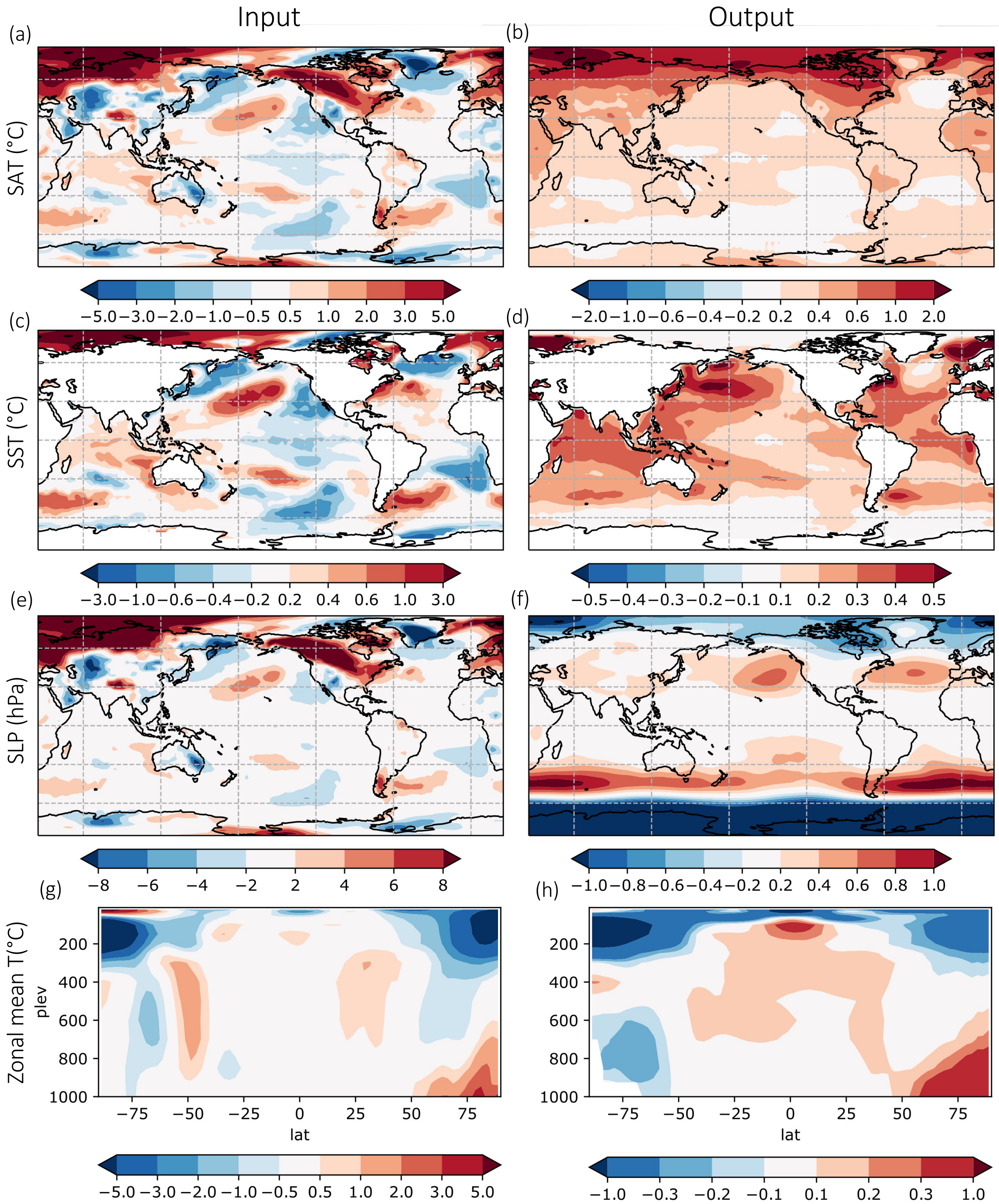

Fig. 2. Example of a randomly selected month, October 2010, for the U-Net inference, using observations and reanalyses as input data: (a) actual surface air temperature anomaly, in °C, used as input and (b) the result of U-Net_tas representing the forced surface air temperature (SAT) anomaly, in °C. (c) and (d) are the same as (a) and (b), but for the sea surface temperature (SST). (e) and (f) are the same as (a) and (b), but for the sea-level pressure (SLP), in hPa. (g) and (h) are the same as (a) and (b), but for the zonal-mean temperature. Note the difference in colorbar scales between left and right panels.

An example of inference is given in Fig. 2 when using fields from observations as inputs. All panels show the anomalies corresponding to a randomly selected month, which is in October 2010. The rights panels are the estimated forced changes obtained for that month. While the input surface air temperature (Fig. 2a) and SST (Fig. 2c) show anomalies of both signs resulting from the internal variability, the forced changes (Figs. 2b and 2d) are smaller in amplitude with a large-scale warming pattern intensified in the Northern Hemisphere and over land. We also note the largest warming over the Arctic. Such a pattern is similar to that found in large ensembles of historical simulations, when the internal variability has been filtered out using an ensemble average (Bonnet et al., 2021; Deser et al., 2020). The pattern is consistent with an amplification of the warming over the Arctic (Previdi et al. 2021) and over land (Lee et al. 2021) as a response to the rising greenhouse gas concentration. The SLP forced anomalies show weak forced anomalies (Fig. 2f). The overall pattern with negative anomalies poleward of 70° in both hemispheres and positive anomalies over 30°S and 50°S in the Southern Hemisphere is reminiscent of the results shown by multi-model means of coupled models (Knutson and Ploshay, 2020), with a poleward shift of the subtropical high-pressure belt. The forced zonal-mean air temperature anomalies estimated (Figs. 2h) are also quite similar to those obtained with coupled models (Lee et al. 2021), with stratospheric cooling and warming located in the Arctic lower troposphere and tropical upper-troposphere. The patterns found show that the estimates obtained are qualitatively consistent with the effect of external forcings found in the training models.

## 3. Skills in separating the forced and internal variability

*a. Temperature and sea-level-pressure*

We first evaluate the skills of the U-Nets in capturing the effect of external forcing of surface air temperature, SST, SLP, and zonal-mean air temperature.

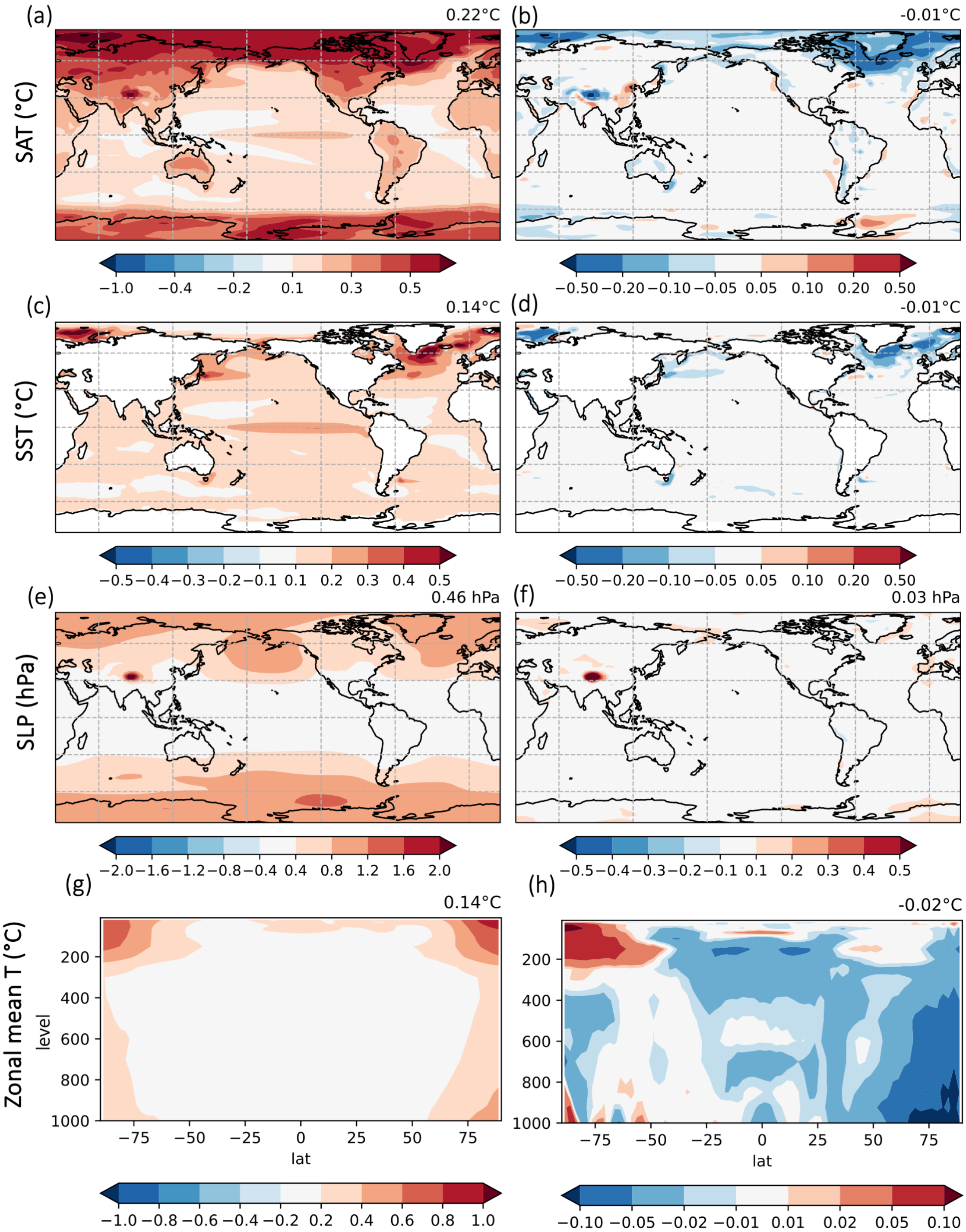

Fig. 3. Root mean squared error (RMSE) and mean error in 2003-2022 for the U-Nets. The error is defined at each time step as the mean difference between the U-Net output and the ensemble mean for the model left out in the validation. (a) RMSE of the surface air temperature (SAT), in °C, and (b) error of the surface air temperature (SAT) in 2003-2022, in °C. (c) and (d) are the same as (a) and (b), but for the sea surface temperature (SST). (e) and (f) are the same as (a) and (b), but for the sea-level pressure (SLP), in hPa. (g) and (h) are the same as (a) and (b), but for the zonal-mean air temperature. The global mean for each map is indicated in the top right of each panel.

We calculate the RMSE based on the period 1950-2022 and, for the left-out model, based on the difference between the U-Net estimate obtained for each member and the ensemble mean. The cross-validation error is obtained by averaging the RMSE obtained from the five training sets, each excluding one model at a time. It is calculated at each grid point at the monthly frequency and is provided in the left panels of Fig. 3. The right panels of Fig. 3 show the difference between the mean estimated forced variability and the ensemble mean using a time average in 2003-2022. This period corresponds to the end of the time series, where the influence of the increasing greenhouse gas concentration is expected to be the largest. The errors obtained can be attributed to the training data, which can be too different from the left-out model. The ensemble mean also includes an error when estimating the forced model response, as the internal variability is only reduced by a factor of $\sqrt{n}$, $n$ being the size of the ensemble. Therefore, the inter-model spread was calculated in Fig. S6 (left column) to quantify the lack of robustness in the dataset. This inter-model spread is calculated by the time-averaged standard deviation across the five models' ensemble means. The internal variability is also calculated by the multi-model mean of the ensemble mean temporal standard deviation, and is shown in Fig. S6 (right column) to estimate where the ensemble mean estimate has the largest errors.

For the surface air temperature and SST, the errors are typically of the order of 0.1 to 0.3°C in the tropics, with 0.2°C to 0.3°C of error over land and the eastern Equatorial Pacific (Fig. 3ac) and errors below 0.1°C over the warm pool and the western Atlantic. The errors in these tropical locations correspond well to the spread among models (Fig. S6ac). The internal variability is also large at these locations, with El Niño Southern Oscillation (ENSO) being associated with SST anomalies over the eastern Equatorial Pacific (Fig. S6bd). The SST errors are also large, with up to 0.4°C in the Atlantic and Pacific Ocean north of 40°N (Fig. 3c), with maximum errors over the western boundary currents and at the sea ice edge. The surface air temperature also shows errors of about 0.4°C over North America, Siberia and Antarctica (Fig. 3a). The surface air temperature over the region covered by sea ice in the Southern and Northern Hemisphere shows the largest error, with up to 0.5°C over the Barents Sea. Again, the low agreement among models explains these errors. The bias in 2003-2022 (Fig 3bd) illustrate a small underestimation of the warming of about 0.1°C located in the North Atlantic and Arctic region, with otherwise biases below 0.05°C in absolute value.

We explored the same diagnostic for each validation model, which reveals similar RMSE patterns, but with a larger RMSE when using CanESM5 in validation (global mean RMSE of 0.29°C) compared to the other models (global mean RMSE of 0.19°C, 0.19°C, 0.18°C and 0.22°C). The underestimation of the warming in 2003-2022 is only obtained when using CanESM5 in validation (mean difference of -0.16°C), while it is almost absent when using the other models in validation (0.02°C, 0.00°C, 0.03°C and 0.03°C). Lastly, all the tests also found much larger loss functions (see Figs. S2, S4 and S5) when using the data of CanESM5 in validation. This corresponds to the transient climate sensitivity of CanESM5, which is larger than that of the other models (Zelinka et al., 2020). Therefore, we speculate that the U-Net struggles to detect a large forced warming that is not included in its training data.

The forced variability estimated by the fourth-order polynomial trend shows the same pattern as Fig. 3, but with a larger RMSE (see Fig. S8ac). When using the fourth-order polynomial trend, the global mean bias in 2003-2022 is much reduced, but the local biases outside of the Arctic region are larger (Fig. S8bd). The errors and biases are also larger than those of Fig. 3 when using the previous U-Net version used in Wills et al. (2026), with an even larger underestimation of the forced warming in 2003-2022 (see Fig. S9abcd), in agreement with the better optimization of the U-Net of this study.

The SLP shows RMSEs of about 1 hPa centered over mid-latitude ocean poleward of 50° in both hemispheres, while tropical regions have errors below 0.4 hPa (Fig. 3e). This corresponds well to regions with a large inter-model spread and with a large internal variability (Fig. S6ef). We can note that the internal variability of the SLP is large in these regions (approximately 5 to 7 hPa), which may alter the estimation of the forced changes. The SLP biases (Fig. 3f) are weak and below 0.1 hPa in 2003-2022. Again, the errors and the local biases in 2003-2022 are found to be larger when using the fourth-order trend estimate (Fig. S8ef) or in the Wills et al., (2026) U-Net version (Fig. S9ef).

The zonal-mean temperature RMSE (Fig. 3g) is largest in the polar stratosphere poleward of 75° in both hemisphere (~0.5 to 0.8°C) and the polar lower troposphere (up to 0.5°C), as the models show a poor agreement and a large internal variability at these locations (Fig. S6gh). The bias during 2003-2022 is locally between 0.02°C and 0.05°C (Fig. 3h), with again an underestimation of the forced signal. Again, the error and the biases can be attributed to the poor estimation of the warming when using the out-of-distribution CanESM5 data in validation. The RMSE are again reduced when compared to the fourth-order trend estimate (Fig. S8gh) or to Wills et al. (2026) U-Net estimates (Fig. S9gh).

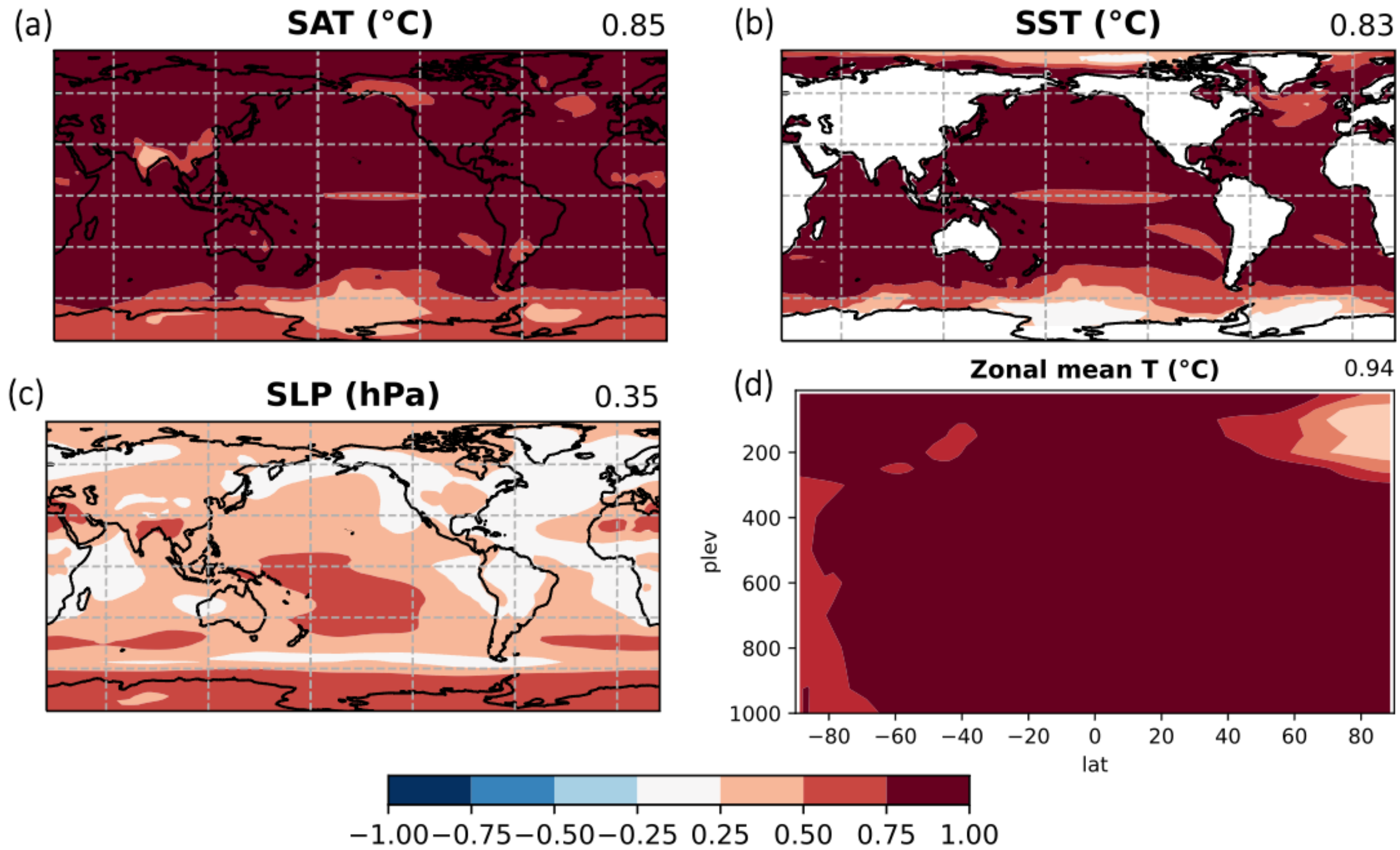

Fig. 4. Mean temporal correlation between the U-Net output and the ensemble mean for all members of the left-out model. The correlation shown is the multi-model mean, using successively the data from one model left out. (a) is for the surface air temperature (SAT), (b), for the sea surface temperature (SST), (c), for the sea-level pressure (SLP) and, (d), for the zonal-mean temperature. The global mean for each map is indicated in the top right of each panel.

We calculate the temporal correlation between the ensemble mean and the estimated forced variability for each member of the left-out model in the validation procedure. Figure 4 shows the mean correlation obtained between the five validation models. The surface air temperature (Fig. 4a) and SST (Fig. 4b) show correlations above 0.75 except locally over the Southern Ocean, where it is between 0.25 and 0.5. The global mean correlation is 0.85 and illustrates that 72% ($R^2$=$0.85^2$=0.72) of the forced variability of surface air temperature is explained. Such temporal correlation is small for the SLP (Fig. 4c), with a value between 0.25 and 0.5, so the forced variability remains poorly estimated. Lastly, the correlation is larger than 0.75 almost everywhere for the zonal-mean temperature (Fig. 4d). The correlation is only reduced in the stratosphere north of 60°N. These correlations are almost everywhere larger than those obtained with the fourth-order polynomial trend (Fig. S10) or with the previous U-Net version (Fig. S11).

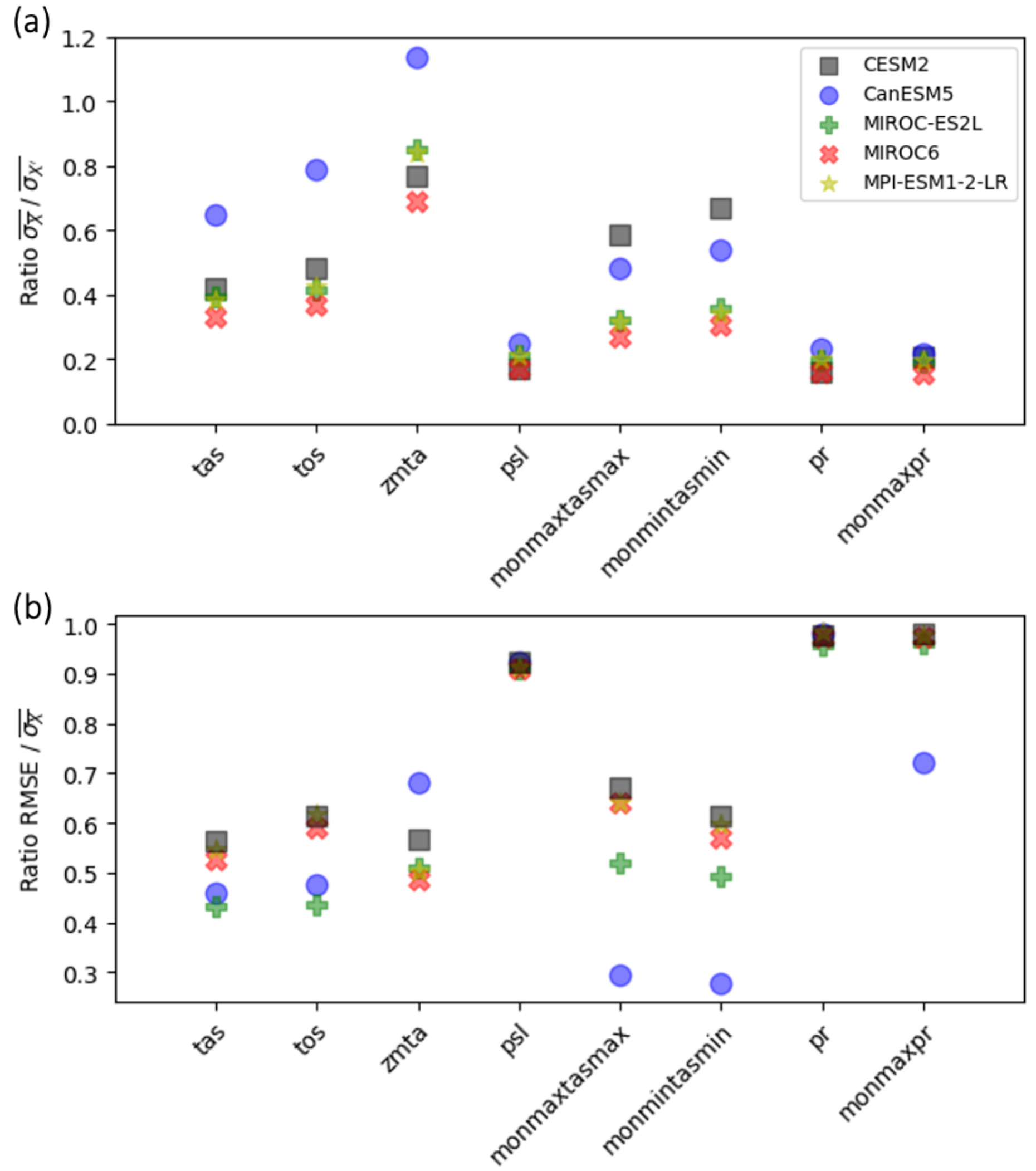

Fig. 5. (a) Ratio between the estimated external and internal variability and (b) between the RMSE of the U-Nets and the external variability. The external variability is given by the global mean of the time standard deviation of the ensemble mean anomalies. The internal variability is given by the global mean ensemble mean of the time standard deviation obtained with the deviations from the ensemble mean. The RMSE designates here the global mean cross-validation error (see Section 2e).

The skills of the U-Net are summarized and discussed in Fig. 5. Panel (a) shows the ratio of external and internal variability. This ratio is given by the global mean temporal standard deviation of the ensemble mean, $\overline{\sigma}_{\overline{X}}$, divided by the global mean ensemble mean temporal standard deviation of the deviations from the ensemble mean, $\overline{\sigma}_{X'}$. Figure 5b also gives the ratio of the global mean RMSE over the external variability, $\overline{\sigma}_{\overline{X}}$. The surface air temperature and the SST show a ratio of approximately 0.5 between the external variability and the larger internal variability. The errors of the U-Net are almost half as large as the external variability for these two variables. The large sensitivity of CanESM5 (blue circle) is evident from the large external-to-internal ratio, but the RMSE relative to the external forcing amplitude is not different from that obtained with the other models when using that model in validation. The SLP has a small ratio of external to internal variability (~0.2), which may explain why the errors are of the same amplitude as the external variability for this variable. The zonal-mean air temperature shows a large external to internal variability ratio of about 0.8, but the errors remain of similar amplitude as the ones obtained with the surface temperature or SST, presumably due to the relatively large inter-model spread for this variable (Fig. S6g).

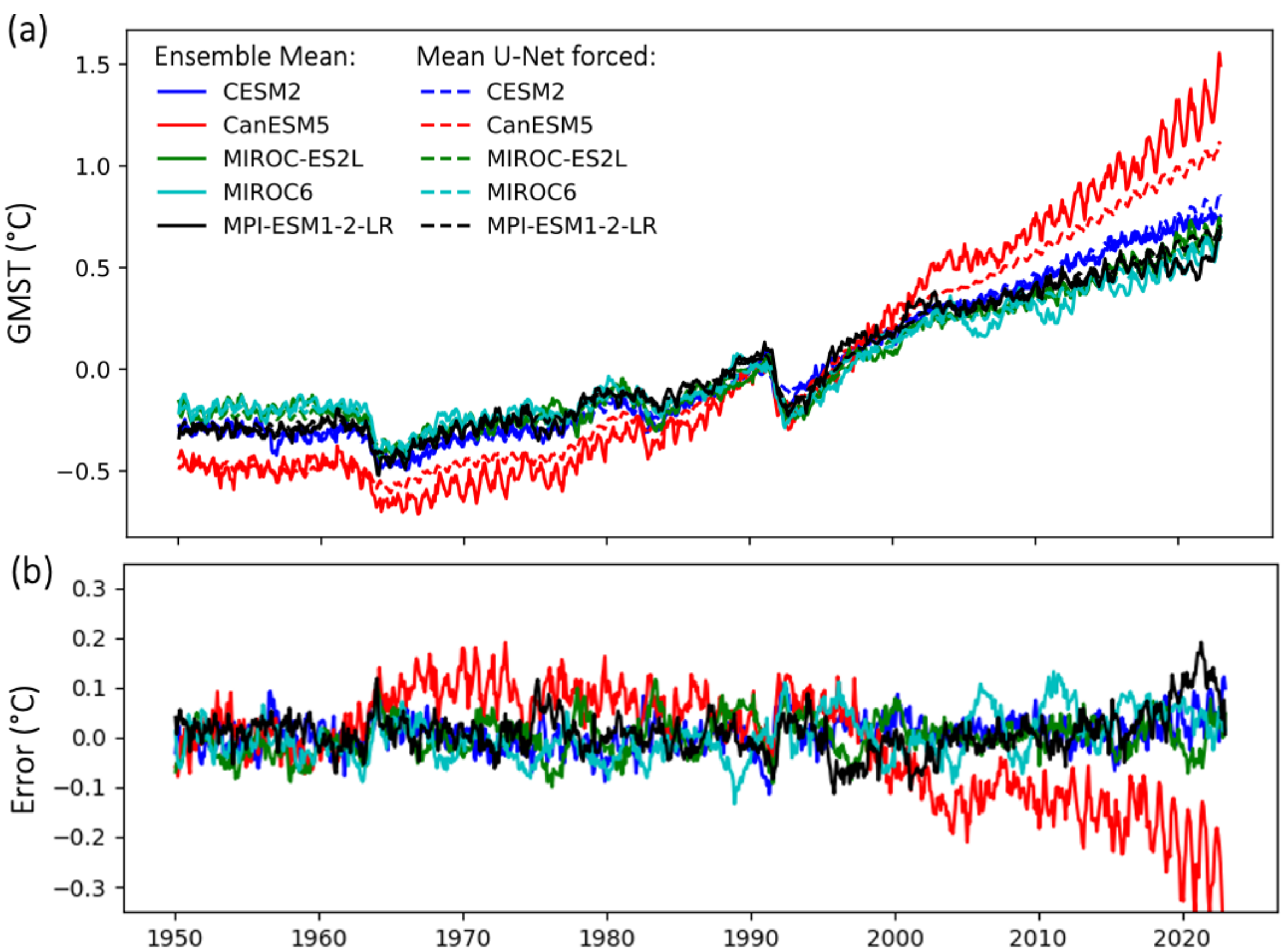

Fig. 6. Time series of the forced global mean surface air temperature (GSAT) anomaly, in °C. (a) Ensemble mean (ground truth, full line) and forced GSAT variability estimated by the U-Net (dashed line) for the model left out in the validation. (b) Mean error, in °C, given by the ensemble mean difference between the U-Net forced GSAT estimate and the ensemble mean for the model left out in the validation.

The time evolution of the errors is investigated for the global surface air temperature (GSAT). We calculate the GSAT from the global mean surface air temperature anomalies, derived from the forced variability captured by the U-Net, and compare it with the actual ensemble mean for each model left out in the validation (Fig. 6). When using CanESM5 as the left-out model (red lines), the estimated forced GSAT (dashed line) is underestimated by 0.1 to 0.3°C in 2005-2022. CanESM5 simulates a warming much larger than the other models because of its large climate sensitivity (Zelinka et al., 2020). When this model is left out, the training data show a smaller warming during the 1950-2022 period, which leads to an underestimation of the warming. Associated with this underestimation, the amplitude of the error increases toward the end of the time series after 2010. Conversely, the forced GSAT variation is well reproduced before 2010, with GSAT errors smaller than 0.1-0.15°C. When using the other models for the validation, we do not find any systematic bias. Some modifications of the seasonal cycle are well reproduced by the forced GSAT estimates. Interestingly, the error in estimating the forced GSAT does not increase a few years after the Agung (1963), El Chichón (1982) or Pinatubo (1991) major volcanic eruptions. This contrasts with the fourth-order trend estimate, for which the error increases both before and after these events (Fig. S12), since the timescale of these anomalies is not captured by a trend.

Next, we investigate the performance of the U-Net for temperature extremes and precipitation.

*b. Temperature extremes and precipitation*

When using TXx or TNn in the U-Net, the RMSE, biases, and correlations (see Figs 7e-h and 8ab) are comparable to those obtained with the monthly surface air temperature. This suggests that the changes of the extreme cold and warm events are similar to that of the mean temperature, with the U-Net performing similarly in detecting the forced changes. Figure 5 shows that TXx and TNn have a ratio of 0.5 between the external and internal variability, which is similar to the surface air temperature, and that their error is also half that of the external variability. However, the ratio of external-to-internal for each model individually shows some disagreement when compared to the surface air temperature, so that processes other than the climate sensitivity, such as the land-atmosphere feedbacks involving soil moisture (Vogel et al., 2018), might explain the inter-model spread.

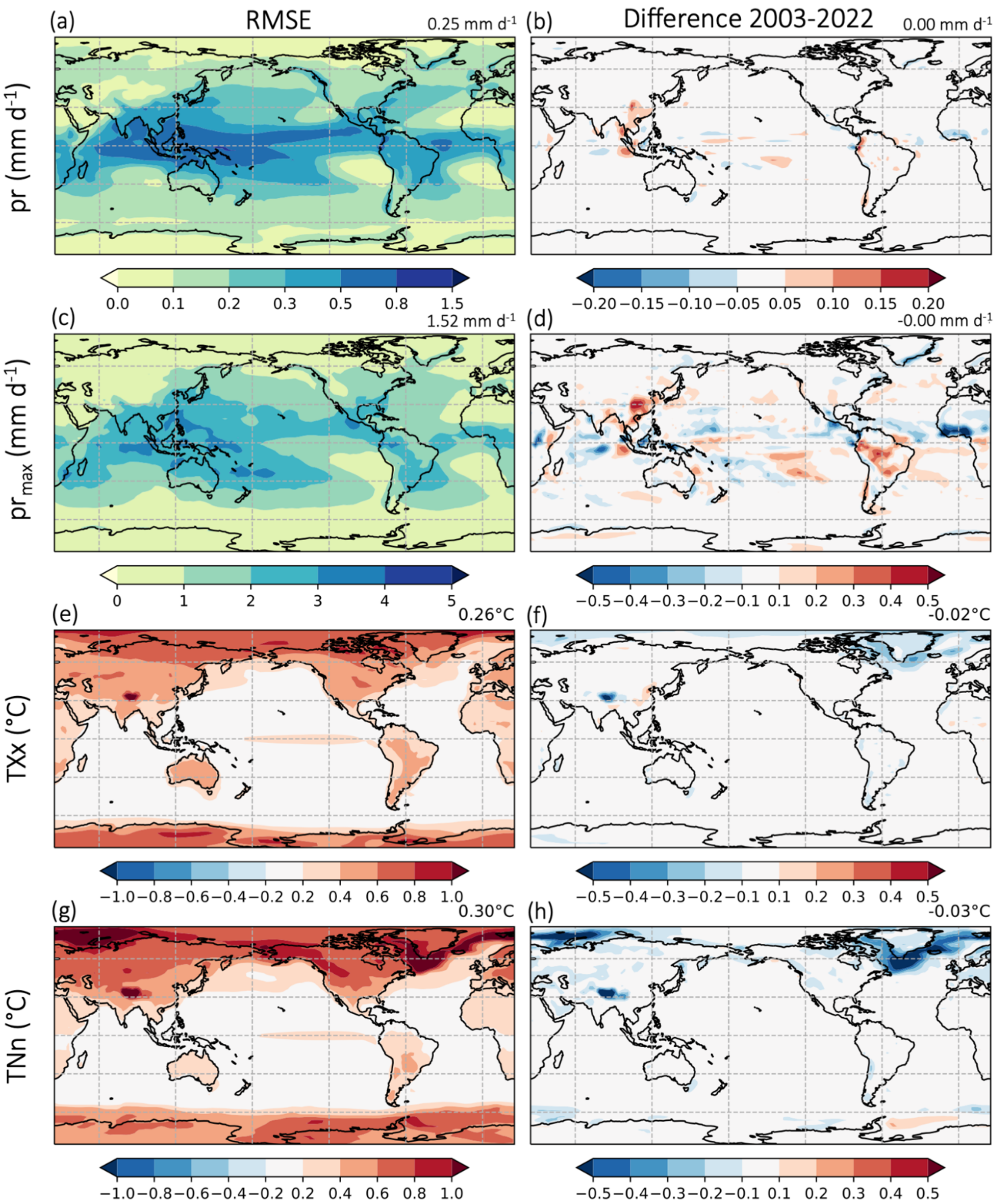

Fig. 7. Same as Figure 3, but for the U-Net dedicated to the (a-b) precipitation, pr, in mm $d^{-1}$, (c-d) monthly maximum of precipitation, $pr_{max}$, in mm $d^{-1}$, (e-f) the monthly maximum of the diurnal maximum air temperature, TXx, in °C, and (g-h) the monthly minimum of the diurnal minimum air temperature, TNn, in °C.

The forced precipitation changes are poorly estimated. The RMSE (Fig. 7a) is of the same magnitude as the mean precipitation, while the correlation (Fig. 8c) is approximately 0.25. The errors in 2003-2022 (Fig. 7b) are mostly over the tropics, where the precipitation is largest and where the displacements of the intertropical convergence zone are not captured. Nevertheless, we note correlations larger than 0.5 over the western Sahel, the Arctic Ocean and the Southern Ocean. The performance of the U-Net for the monthly maximum precipitation shows similar characteristics. The forced variability of precipitation is much smaller than the internal variability (ratio of 0.2 in Fig. 5), and the errors have the same amplitude as the external variability, indicating that the errors are much smaller in amplitude than the internal variability.

For the precipitation or the temperature extremes, the fourth-order polynomial trend results in larger errors (compare Fig. 7 and Fig. S13) and smaller correlations (compare Fig. 8 and Fig. S14) than the U-Net.

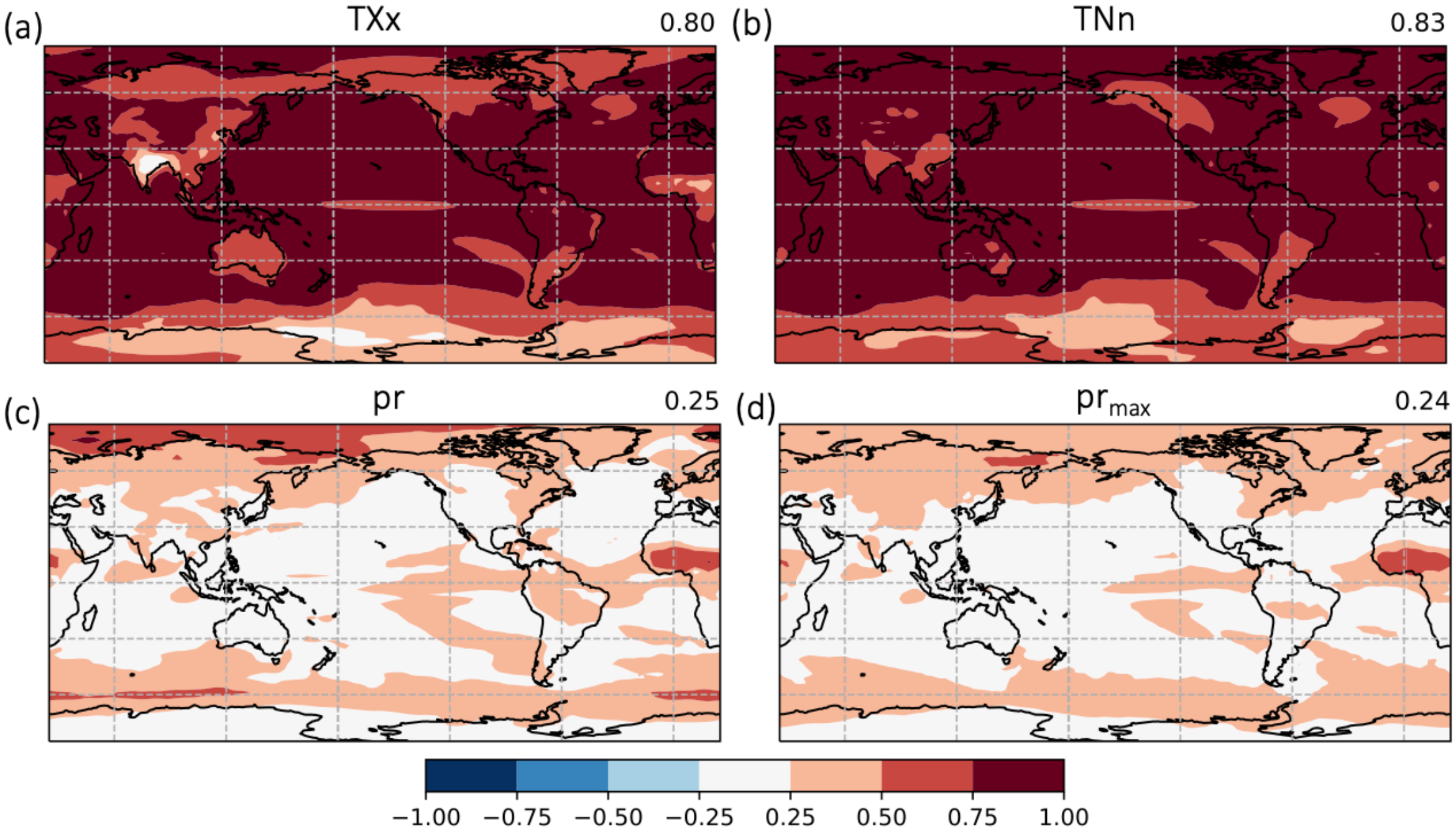

Fig. 8. Same as Figure 4, but for the U-Net dedicated to (a) the maximum of the diurnal maximum air temperature, TXx, (b) the minimum of the diurnal minimum air temperature, TNn, (c) precipitation, and (d) monthly maximum of precipitation, $pr_{max}$.

*c. Climate indices*

The 1979-2022 trends in several climate indices are calculated both for the forced variability captured by the U-Net, and for the actual ensemble mean for each model left out in the validation. We first focus on the yearly GSAT trend. We also examine the AMV, defined as the SST averaged in 0°N-60°N over the North Atlantic, low-pass filtered with a 10-yr cutoff period using a $4^{th}$-order Butterworth filter. The Arctic Amplification (AA), is defined following Sweeney et al. (2023) as the surface air temperature trend poleward of 70°N divided by the GSAT trend. ENSO is investigated with the Niño3.4 index: the SST averaged in 5°N-5°S 170°W-120°W. The tropical upper tropospheric temperature is also analyzed, calculated as the zonal-mean temperature at 200 hPa averaged in 20°N-20°S (Mitchell et al., 2020). The station-based Southern Annular Mode (SAM; Gong and Wang, 1999) is based on the difference in the normalized SLP obtained at six stations located at 40°S and 65°S, respectively. In addition, we calculate the European land TXx in June-July-August (JJA) in 35°N-60°N 11°W-36°E. Finally, mean precipitation is calculated over the Western Sahel 10°N-20°N 20°W-20°E during JJA. Western Sahel is chosen here to illustrate the skill of the U-Net over monsoon regions, as it performs well in this region (see Section 3b).

The error is estimated as two standard deviations, calculated across the U-Net forced estimates obtained from the members of the left-out model. The error in estimating the trend of the ensemble mean is calculated from two standard deviation of trends across ensemble members of the left-out models divided by $\sqrt{n}$, $n$ being the size the left-out model ensemble.

For all the indices based on temperature, the forced trends are generally well captured by the U-Net (Fig. 9). However, when CanESM5 is used as the validation dataset, the U-Net systematically underestimates the forced changes, likely because the training data do not include the large forced variability simulated by this model. The U-Net does not generalize to warming levels not included in the training. Consequently, the GSAT, AMV, Niño 3.4, and the tropical upper tropospheric temperature trends are underestimated when CanESM5 is left out. In contrast, when using MPI-ESM-1-2-LR for the validation, the U-Net slightly overestimates the GSAT and Niño3.4 trends. The trends of the European summer TXx are similarly well captured by the U-Net, with a small underestimation for MIROC-ES2L. The changes in the Arctic Amplification are not well captured by the U-Net, as the U-Net finds values between 2.6 and 3.2, while the models show Arctic amplification between 2.2 and 3.6. Therefore, the U-Net estimates an Arctic Amplification consistent with the multi-model mean of the training data used. The SAM trend is well estimated by the U-Net, except when using MIROC6 in validation. Lastly, the U-Net for the precipitation over the Western Sahel estimates a trend consistent with the multi-model mean and fails to reproduce the more important precipitation trend when using MIROC6 or CanESM5 in validation. If the same analysis is repeated by estimating the forced signal with the fourth-order polynomial trend, we note that there is almost no systematic bias obtained for all indices, but the error is then larger for almost all indices (see the large error bars in Fig. S16).

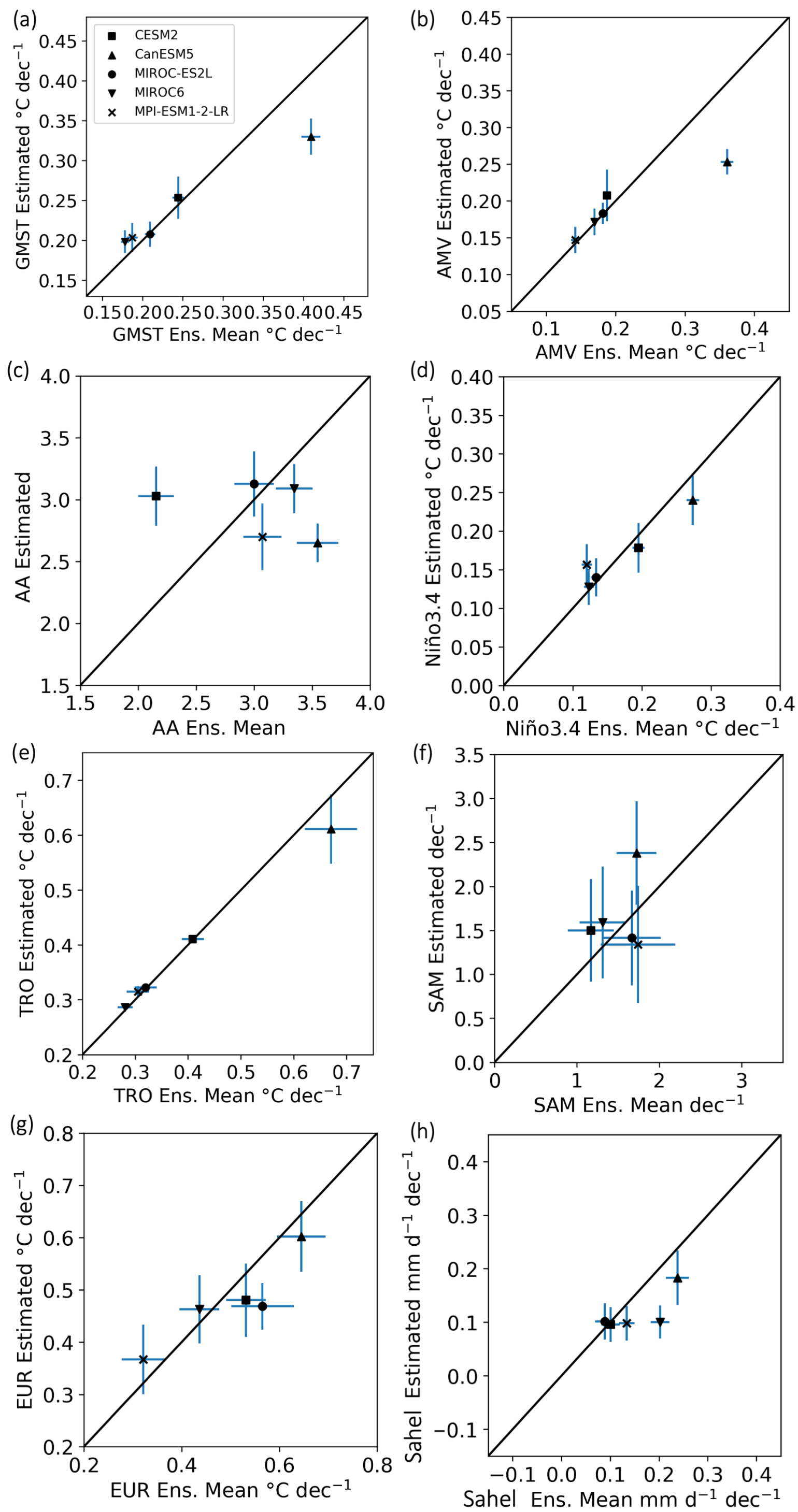

Fig. 9. Scatter plot of the forced 1979-2022 trends estimated versus the ground truth (ensemble mean) 1979-2022 trend for the model left out in the validation for (a) global surface air temperature (GSAT); (b) Atlantic Multidecadal Variability (AMV); (c) Arctic Amplification (AA); (d) Nino3.4 (e) zonal-mean temperature at 200-hPa in tropics (TRO); (f) southern annular mode (SAM); (g) European Summer maximum temperature (EUR, in JJA) (h) Sahel precipitation in JJA. Vertical error bars indicate two standard deviations, calculated across the ensemble members of the model excluded in the training. Horizontal error bars indicate two standard deviations across the ensemble members divided by square root of the ensemble size.

In summary, the forced trends of most indices are captured by the U-Net with an error smaller than those based on estimating the trend. However, common to most machine learning techniques, the U-Net fails to capture forced changes not represented in its training data.

## 4. Inference with observations

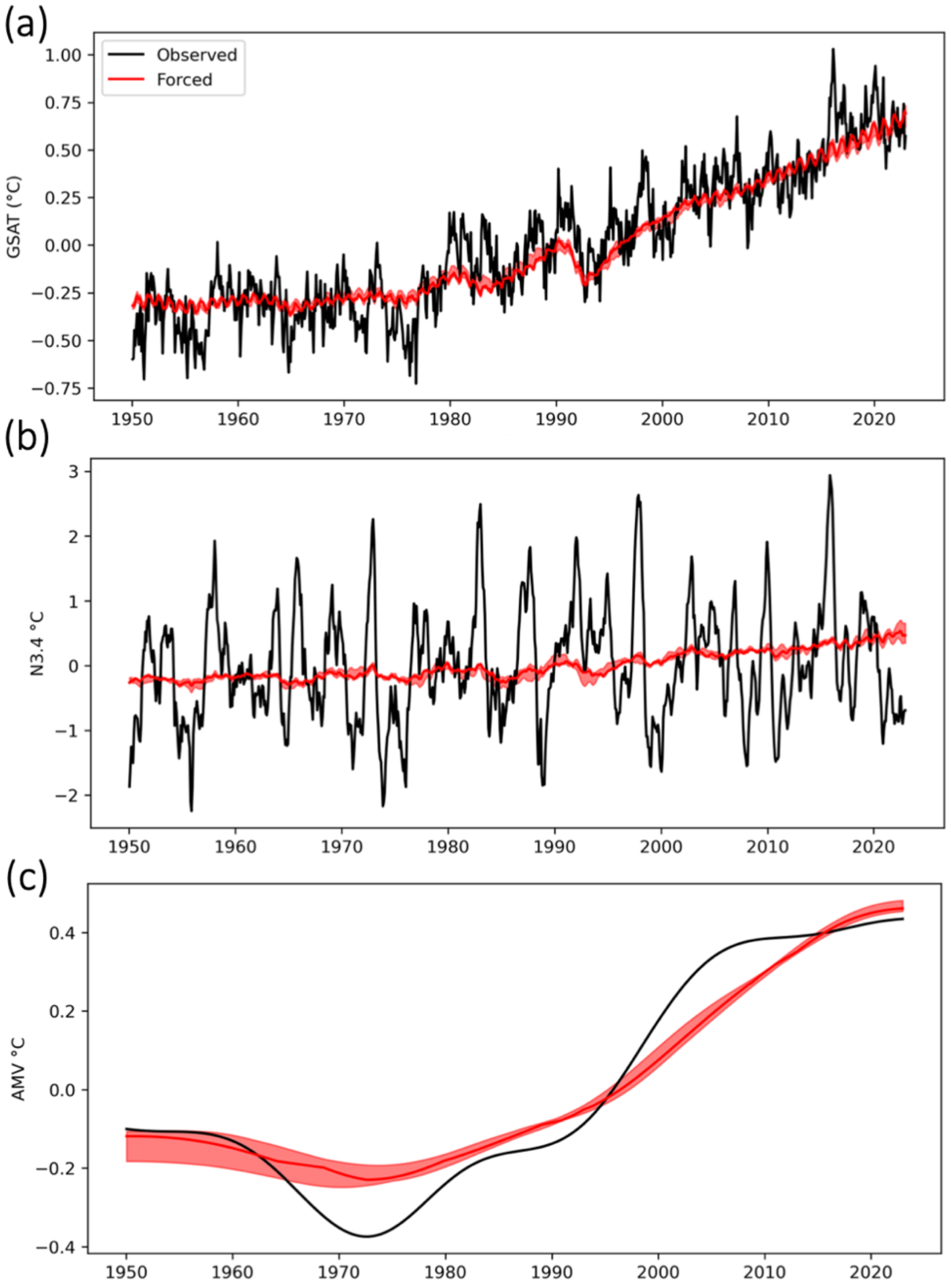

Fig. 10. (a) Time series of the global mean surface air temperature (GSAT) anomaly, in °C, provided as input (black) and the estimated forced changes (red). The red shades indicate the range between the minimum and maximum values of the five estimates, obtained using the U-Nets each trained with one model excluded (see Section 2f). b) and c) are the same as a), but for Niño3.4 and the AMV.

We examine the U-Net inference using observational data as input. The U-Net estimate of the forced GSAT variability (red line in Fig. 10a) follows the observed long-term evolution of the GSAT (black line), with a stable climate in 1950-1975, and a steady trend from 2000 to 2022. Two cool events lasting 4 or 5 years are estimated around 1982 or 1991, which is consistent with the effect of volcanic aerosol injection following the El Chichón and Pinatubo eruptions. The forced estimate captures some changes in the GSAT seasonal cycle, showing stronger warming during Northern Hemisphere winter. The spread among the five estimates is small (see red shades in Fig. 10a). For Niño 3.4, the forced component shows a weak positive trend in all five estimates (Fig. 10b). The low-pass filtered Atlantic SST (Fig. 10c) due to the forced component closely follows the observed variability, suggesting that most of the AMV fluctuations are externally forced, as found by Murphy et al. (2017). Nevertheless, we note that the negative AMV phase in the 1970's and the positive AMV phase in the 2000's are estimated to result from internal variability, with the forced AMV instead showing little change in 1950-1980 followed by a smooth increasing trend.

These results can be interpreted with the trends of the climate indices calculated in the previous section, in addition to the internal variability and multi-model ensemble mean of these trends captured by the five large ensembles (see Table 2). The internal variability is calculated from the multi-model mean of the ensemble standard deviation for the trend in indices. We also apply a simple consistency test: under the null hypothesis that the residual of the forced variability is consistent with the internal variability of climate models, the observed trend should be included in the interval $I$:

$$I = [T_{\mathrm{UNet}} - u_{0.05}\sqrt{\sigma_{\mathrm{UNet}}{}^2 + \sigma_{\mathrm{Int}}{}^2}, T_{\mathrm{UNet}} + u_{0.05}\sqrt{\sigma_{\mathrm{UNet}}{}^2 + \sigma_{\mathrm{Int}}{}^2}] \qquad (1)$$

where $T_{\mathrm{UNet}}$ is the estimated forced trend by the U-Net. $\sigma_{\mathrm{UNet}}$ is the standard deviation across the five U-Net estimates each trained with one model excluded, sampling the effect of using different training data. $\sigma_{\mathrm{Int}}$ quantifies the internal variability with the multi-model mean standard deviation among ensemble member. $u_{0.05}$ is the two-tailed 5% critical value of the standard normal distribution.

|  | Observation | U-Net | Internal variability | Multi-model mean | Consistency test |
|---|---|---|---|---|---|
| GSAT (°C dec$^{-1}$) | 0.190 | 0.208 $\pm$ 0.018 | 0.043 | 0.246 | Not rejected |
| Niño3.4 (°C dec$^{-1}$) | -0.041 | 0.143 $\pm$ 0.038 | 0.126 | 0.169 | Rejected |
| AMV (°C dec$^{-1}$) | 0.190 | 0.170 $\pm$ 0.006 | 0.064 | 0.208 | Not rejected |
| AA | 3.81 | 2.91 $\pm$ 0.21 | 0.986 | 3.02 | Not rejected |
| TRO (°C dec$^{-1}$) | 0.182 | 0.230 $\pm$ 0.027 | 0.004 | 0.397 | Rejected |
| SAM (dec$^{-1}$) | 0.214 | 0.142 $\pm$ 0.048 | 1.916 | 1.523 | Not rejected |
| EUR (°C dec$^{-1}$) | 0.826 | 0.583 $\pm$ 0.111 | 0.283 | 0.500 | Not rejected |
| Sahel (mm d$^{-1}$ dec$^{-1}$) | -0.022 | 0.048 $\pm$ 0.017 | 0.117 | 0.153 | Not rejected |

Table 2. Trends from 1979-2022 of climates indices for: (first column) observations (see section 2 for details), (second column) the forced variability of observation as estimated by the U-Net with the uncertainty provided by two standard deviation of the five U-Net estimates each trained with one model excluded, (third column) the multi-model mean internal variability estimated by two ensembles standard deviations; (fourth column) the multi-model mean ensemble mean and (last column) consistency test at the 5% significance level (see text for details).

For the GSAT trend, the climate models simulate a trend larger than that observed (0.246 °C dec$^{-1}$ vs 0.190 °C dec$^{-1}$), and the U-Net estimates that most of this trend is forced, as the observed trend is within the spread provided by the five U-Net (0.208 $\pm$ 0.018 °C dec$^{-1}$). For the AMV, the observed trend (0.190 °C dec$^{-1}$) is larger the U-Net estimated forced trend (0.170 $\pm$ 0.006 °C dec$^{-1}$), and internal variability ($2\sigma_{\text{Int}}$ = 0.064 °C dec$^{-1}$) can explain this difference. For the trend in Niño3.4, the observed trend is negative (-0.041 °C dec$^{-1}$), while the training models found a positive trend (0.169 °C dec$^{-1}$ for the multi-model trend). The U-Net estimates a forced trend of 0.143 $\pm$ 0.038 °C dec$^{-1}$, and the consistency test is rejected as the internal variability or the spread in training data cannot explain the difference of the observed and U-Net estimate. This emphasizes the limits of the method presented here, which is based on model data and cannot fully explain the observational changes. Arctic amplification and Summer European maximum temperature are two cases where the observed changes are larger than the multi-model mean changes, so the U-Net forced trend is also lower than observations. But in these two cases, the internal variability can explain the differences, and the consistency test is not rejected. Lastly, the western Sahel summer rainfall or the SAM observational trends disagree between models and observations, but in these cases the internal variability is large and can explain the disagreement. All of these statements of consistency between the U-Net estimate and observations pertain to the 1979-2022 trends and may vary with the trend period.

## 5. Conclusions

A U-Net is trained using climate model data to separate the internal and forced components of climate variability in the 1950-2022 period. We apply a U-Net, which is a standard architecture in image processing and computer vision, with 3D kernels to detect spatio-temporal features associated with external forcing, as in Bône et al. (2024). Such a framework does not assume that the spatial patterns are invariant for the forced or internal variability. The U-Net is trained using five single-model initial-condition large ensembles of climate model simulations. We use a supervised setting, using in the training an ensemble member of one model as an input and the ensemble mean of the same model as an output. In addition, the U-Net uses multiple fields with surface air temperature, SST, SLP and the zonal-mean air temperature, as these variables have been regularly observed since 1950. The evaluation of the U-Net results was presented in Wills et al. (2026) for a preliminary version of the U-Net. However, we present the method in detail here, document the cross-validation skill and show the results of U-Nets with improved optimization, resulting in improved skills.

The validation error is established using a cross-validation, excluding successively the model data from one of the five models of the training data. The U-Net method shows validation errors of approximately 0.1°C to 0.4°C for the monthly surface air temperature fields, with the largest error over Arctic or land regions. This error is relatively small, as the forced anomalies are approximately twice as large. The errors are mainly due to one model in the training data (i.e. CanESM5) that has a large climate sensitivity. The U-Net cannot capture the large warming of CanESM5 when trained with the other four models. It is therefore key that the U-Net is trained with data representative of the changes investigated. Interestingly, the U-Net estimates well the climate anomalies following major volcanic eruptions, which can be notoriously difficult for other methods (Frankignoul et al. 2017). The relatively low errors in surface air temperature translate into a good validation skill at estimating the forced changes of various climate indices such as Niño3.4 or the AMV. In contrast, the U-Net struggles to estimate the forced changes of the SLP, with an RMSE of 0.8 hPa to 1.2 hPa in extratropical regions, for forced changes that are of similar amplitude. This result is attributed to the small signal-to-noise ratio for the forced change of this variable, and to the weak multi-model agreement in the training data regarding the forced changes. The U-Net is also applied to precipitation changes, but the method then also performs poorly in separating the forced and internal variability, although results suggest some skills over the western Sahel or the Arctic region. In all cases, the U-Net gives better estimates of the forced changes than a fourth-order polynomial trend. However, the cross-validation error presented here is a lower bound of the actual error, as the hyperparameters are chosen to minimize the validation.

A first assessment of the forced climate variability inferred from observations indicates that internal variability plays a major role in the 1979-2022 changes in Niño 3.4, European summer temperature and Arctic Amplification, while the AMV appears to be mainly driven by the forced component, which corroborates previous investigations (Sweeney et al., 2023; Murphy et al., 2017 and Freveletti et al., 2026). However, the method presented may have limitations when applied to data that differs substantially from those used for training. Regarding Niño 3.4, the observed trend is negative (La Niña-like) and is outside of the spread given by the internal variability of models plus the U-Net forced change (El Niño-like). Thus, the changes observed are not in agreement with the model training data and the generalization of the U-Net might be limited regarding this aspect. This issue is linked to the limitation of climate model data in reproducing the long-term trend in the Equatorial Pacific Ocean (Kociuba and Power, 2015; Watanabe et al., 2021, Wills et al. 2022; Watanabe et al. 2024). A similar disagreement between observation and model data is also obtained for the upper tropospheric tropical temperature, which warms less in observations than in models. This is presumably linked with the poor agreement between observation and models regarding the Equatorial Pacific long-term changes. Lastly, we focus on the summer European maximum daily temperature, for which the observed trend is larger than the U-Net estimated forced trend, as in Vautard et al. (2023), but in this case the internal variability can explain the difference.

Given the nature of the method presented, we speculate that the results could be improved by increasing the size and diversity of the training dataset, but this remains to be investigated. At least eighteen single-model initial-condition large ensembles are now available (Maher et al. 2025), which could supply additional data for the training. It is likely helpful for the generalization of the U-Net to include models with a wide range of climate sensitivities, or to include models prone to simulate a La-Niña-like trend (Liu et al., 2025). Similarly, we use the ensemble mean as an output to train the U-Net to recognize the forced anomalies. Other estimates of the forced anomalies from large ensembles, for example using a signal-to-noise maximizing pattern filter (Wills et al., 2020), could be tested in the output of the training data. Although several tests have been carried out to optimize the U-Net and refine its architecture, there are still several modifications that can be tested. This is especially the case for variables other than the surface air temperature, as the limited GPU time has reduced the search for optimal hyperparameters. Lastly, diffusion models or flow matching are largely applied in computer vision (Ho et al., 2020). Using the U-Net presented here might be considered to extract the forced variability and build a model that would be able to simulate the internal climate variability using neural networks derived from AI-weather forecast models such as AIFS (Lang et al., 2024), ACE2 (Watt-Meyer et al., 2025) or ArchesClimate (Clyne et al., 2025).

*Acknowledgments.*

We acknowledge the support of the EUR IPSL Climate Graduate School project managed by the ANR under the "Investissements d'avenir" programme with the reference ANR-11-IDEX-0004-17-EURE-0006. This work was performed using HPC resources from GENCI-IDRIS AD011013295. G.G. was funded by the French National program SUN. R.C.J.W was supported by the Swiss National Science Foundation (Award PCEFP2 203376). We also acknowledge funding from the Swiss National Science Foundation Scientific Exchanges program (Award IZSEZ0 220740), which supported a scientific exchange between authors as part of the ForceSMIP hackathon.

*Data Availability Statement.*

The CMIP6 source data are available via ESGF, and the processed large ensemble data used in ForceSMIP have recently been made available by Maher et al. (2025). ERA5 data are available from https://cds.climate.copernicus.eu/datasets. ERSSTv5 data is available from https://psl.noaa.gov/data/gridded/data.noaa.ersst.v5.html. The ForceSMIP Tier 1 data, i.e., the raw data, ensemble means, and estimated forced responses for each variable and each evaluation member is available on Zenodo (https://doi.org/10.5281/zenodo.15577519; Wills et al. (2025)). The code for the U-Net, as well as the other methods used in ForceSMIP and published in Wills et al. (2026), is available via Github (https://github.com/ForceSMIP/tier1-methods). The code of the optimized U-Net version is provided via the Github repository

https://gitlab.in2p3.fr/guillaume.gastineau/unet3d-internal-and-forced-climate-variability and the updated forced response for observations is available at https://doi.org/10.5281/zenodo.21982748.